\documentclass[1p,authoryear]{elsarticle}

\makeatletter
\def\ps@pprintTitle{%
 \let\@oddhead\@empty
 \let\@evenhead\@empty
 \def\@oddfoot{\hfill \footnotesize \emph{\today}}%
 \let\@evenfoot\@oddfoot}
\makeatother

\usepackage[utf8]{inputenc}
\usepackage{graphicx}

\usepackage{algorithm,algorithmicx,algpseudocode}
\usepackage{amsmath,amssymb,amsfonts,amsbsy,amsthm,booktabs,epsfig,graphicx,rotating}

\usepackage[capitalise]{cleveref}
\usepackage[caption=false]{subfig}

\usepackage{natbib} 
\journal{IJoG}

\begin{document}

\begin{frontmatter}

\title{Newcomb-Benford's law as a fast ersatz of discrepancy measures}

\author[1]{Pamphile T. Roy\corref{cor1}}
\ead{roy.pamphile@gmail.com}

\cortext[cor1]{Corresponding author}

\address[1]{iTranslate GmbH, Gadollaplatz 1, 8010, Graz, Austria}

\begin{abstract}

Thanks to the increasing availability in computing power, high-dimensional engineering problems seem to be at reach. But the curse of dimensionality will always prevent us to try out extensively all the hypotheses. There is a vast literature on efficient methods to construct a Design of Experiments (DoE) such as low discrepancy sequences and optimized designs. Classically, the performance of these methods is assessed using a discrepancy metric. Having a fast discrepancy measure is of prime importance if ones want to optimize a design. This work proposes a new methodology to assess the quality of a random sampling by using a flavor of Newcomb-Benford's law. The performance of the new metric is compared to classical discrepancy measures and showed to offer similar information at a fraction of the computational cost of traditional discrepancy measures.


\end{abstract}


\end{frontmatter}


\section{Introduction}


Newcomb-Benford's law---see \citep{Newcomb1881,Benford1938}---states that the occurrence of the first digit of a sequence follows a logarithmic distribution

\begin{align}
	\Pr(d) = \log_{10}(1 + 1/d),
\end{align}
\noindent with $d \in \mathbb{N}$. One paramount condition is that the sequence should span over multiple orders of magnitude to be true. The law can be extended to next significant digits---see~\citep{Hill1995a}. Starting from the 4th digit, the probability is almost equi-distributed to all 10 digits with $\Pr(d) \simeq 10\%$. Note that the frequency of the digits is not independent. Hence, we cannot test for multiple digits occurrence at the same time without decreasing the power of the tests. Instead, the law is still valid if we are looking at the joint distribution of the leading digits. $1.23$ can be seen as either $1$, $12$ or $123$. Doing so increases the power, but marginally (5\%)---see~\citep{Joenssen2013}. The loss of information due to rounding explains this behaviour.

The Newcomb-Benford's law has seen many applications ranging from fraud detection, to image analysis. Hill gives an explanation in~\citep{Hill1995a}: if random samples are generated from distributions selected at random, then the significant digits of the sample converge to the logarithmic distribution. Then, one can naturally wonder:\\

Can Newcomb-Benford's law be used to assess the randomness of samples?\\

A sample corresponds to a given set of input parameters $x_k$ with $k \in [1, \dots , d]$, and $d$ is the number of dimensions. The set of $N_s$ samples is noted as $\mathbf{X}^{N_s}_d$ and also called a Design of Experiments (DoE). DoE have numerous uses from numerical integration to experimental designs~\citep{Sacks1989}. As the dimension grows, the volume of the hypercube increases exponentially. It quickly becomes intractable to completely fill the space. 

Different metrics are used to characterize the space filling of samples. Mainly, there are geometrical and uniformity criteria. They respectively measure a distance between all points and measure how the points position deviates from the uniform distribution~\citep{Fang2006,Androulakis2016}. The latter is referred to as the discrepancy. Hence, using low-discrepancy methods such as Sobol'~\citep{Sobol1967} is a common practice. There are various ways to calculate the discrepancy, but they all share a common issue: their numerical complexity. The well used $C^2$-discrepancy is $\mathcal{O}(N_s^2\times d)$. When the measure is used in an optimization loop, the numerical cost can become quickly intractable.

This work proposes a new method to assess the space-filling of a sample which alleviates the computational complexity incurred with classical discrepancy measures. Starting from the sample to characterize, the values are transformed to the logarithmic law. The set is then compared to Newcomb-Benford's law which gives a metric. These operations are very simple to perform and scalable.

The paper is organized as follows. \Cref{sec:method} describes how to construct a discrepancy metric using Newcomb-Benford's law. \Cref{sec:results} demonstrates the performance of the method with respect to classical method used to assess the quality of a DoE.
Finally, conclusions and future work are drawn in~\cref{sec:conclusion}.

\section{Presentation of the Newcomb-Benford discrepancy} \label{sec:method}

The basis of the method is to compute the deviation from Newcomb-Benford's law. First, the sample is scaled from the unit hypercube to the logarithmic law. Each sample's first significant digit is then counted which leads to individual probabilities of occurence per digit. Goodness of fit is then used to compare this data with the logarithmic law which gives a metric. \Cref{alg:nbd} gives an overview of the process. In the following, it is referred to as the Newcomb-Benford discrepancy (NBD). 

\begin{algorithm}
  \caption{Newcomb Benford's discrepancy}
  \label{alg:nbd}
  \begin{algorithmic}[1]
  \Require $\mathbf{X}^{N_s}_d$
  \Comment{Start from a sample $\mathbf{X}^{N_s}_d$ composed of $N_s$ samples in dimension $d$}
  
  \State $\mathbf{X}^{N_s}_d \gets 10^{\mathbf{X}^{N_s}_d}$ 
  \State $\mathbf{X}^{N_s}_d \gets$ Clip values of $\mathbf{X}^{N_s}_d$ between 1, 9 and round down
  \State $\Pr(d) \gets$ Count number of occurrences of each digit and divid by $N_s \times d$
  \State $f \gets$ RMSE between Newcomb-Benford's law and $\Pr(d)$

  \end{algorithmic}
\end{algorithm}

Goodness of fit is computed using Cramér-von Mises $U_d^2$. It was shown in \citep{Lesperance2016} to be a robust method to check the conformance with Newcomb-Benford's law. Its discrete form corresponds to the Root Mean Square Error (RMSE) as
\begin{align}
\epsilon = \left( \frac{1}{9} \displaystyle \sum_{i=1}^{9} \left(\Pr(d_i) - \Pr_{N_s\times d} (d_i)\right)^2 \right)^{1/2},
\end{align}
\noindent with $d_i$ the digit to consider. The metric uses a flattened array---of size $N_s \times d$---so that all dimensions are taken into account: it is called the Newcomb-Benford discrepancy (NBD). This metric has a computational complexity of $\mathcal{O}(N_s \times d)$. Which is, not only a great improvement over the commonly used $C^2$-discrepancy, but the operations are also arguably simpler than classical discrepancy calculations.

\begin{figure}[!h]               
\centering

\includegraphics[width=0.7\linewidth,height=\textheight,keepaspectratio]{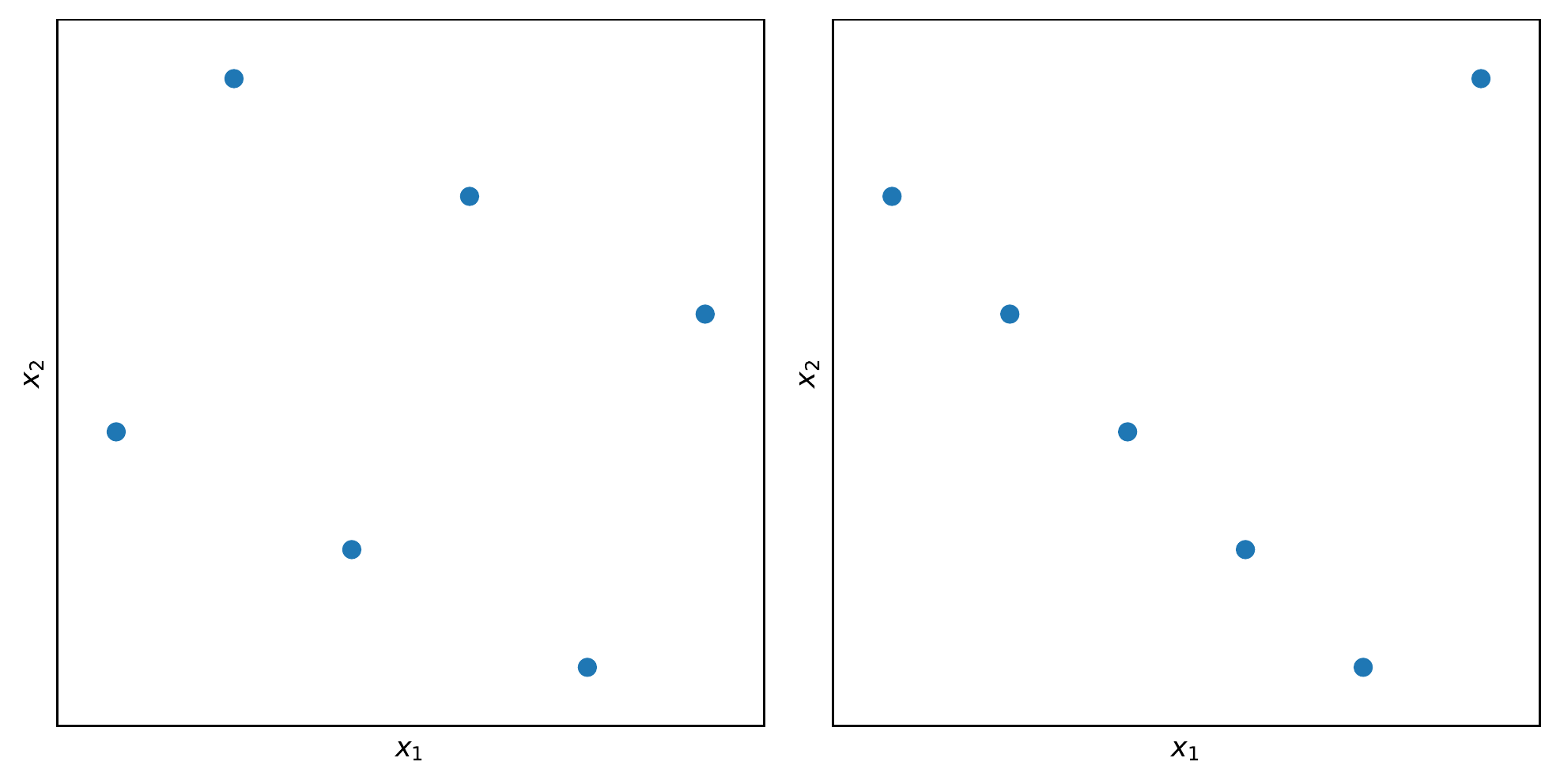}

\caption{LHS design with permuted coordinates.}
\label{fig:lhs}
\end{figure}

One clear drawback when considering all dimensions at once is that the metric is invariant to coordinate permutations. When evaluating LHS designs, this could be an issue as a common optimization scheme consists in permuting the coordinates. \Cref{fig:lhs} shows two different LHS designs which are constructed using a permutation of digits from $[1, 2, 3, 4, 5, 6]$ in 2-dimensions. The design of the left is clearly superior in terms of space coverage. The $C^2$-discrepancy is respectively $0.0081$ and $0.0105$. Still, the computed NBD would be the same as the set of digits would strictly be identical.

To mitigate this issue, I propose to consider all 2-dimensional subprojections of the space---without the diagonal combinations. Hence, another metric can be computed based on the 2-dimensional joint distribution of the coordinates. Considering higher order joint distribution would be intractable as there are $10^d-1$ possible digits to calculate the probabilities from. Hence, considering 2-dimensional projections, the computational complexity becomes $\mathcal{O}(N_s\times (d^2-d))$. It is still an improvement over classical discrepancies as we generally have $N_s >> d$.

\section{Analysis of Newcomb-Benford discrepancy} \label{sec:results}

In the following, crude Monte-Carlo (MC) sampling and the low discrepancy sequences of Sobol' are used to demonstrate the performance of the proposed method. Throughout the literature, and noticeably in~\citep{Kucherenko2015}, Sobol' proved to be superior in every way to MC. Hence, it is expected to show a smaller NBD measure for a given number of samples. Furthermore, the operations have been replicated 99 times to have converged statistics.

\begin{figure}[!h]               
\centering
\subfloat[MC]{
\includegraphics[width=0.5\linewidth,height=\textheight,keepaspectratio]{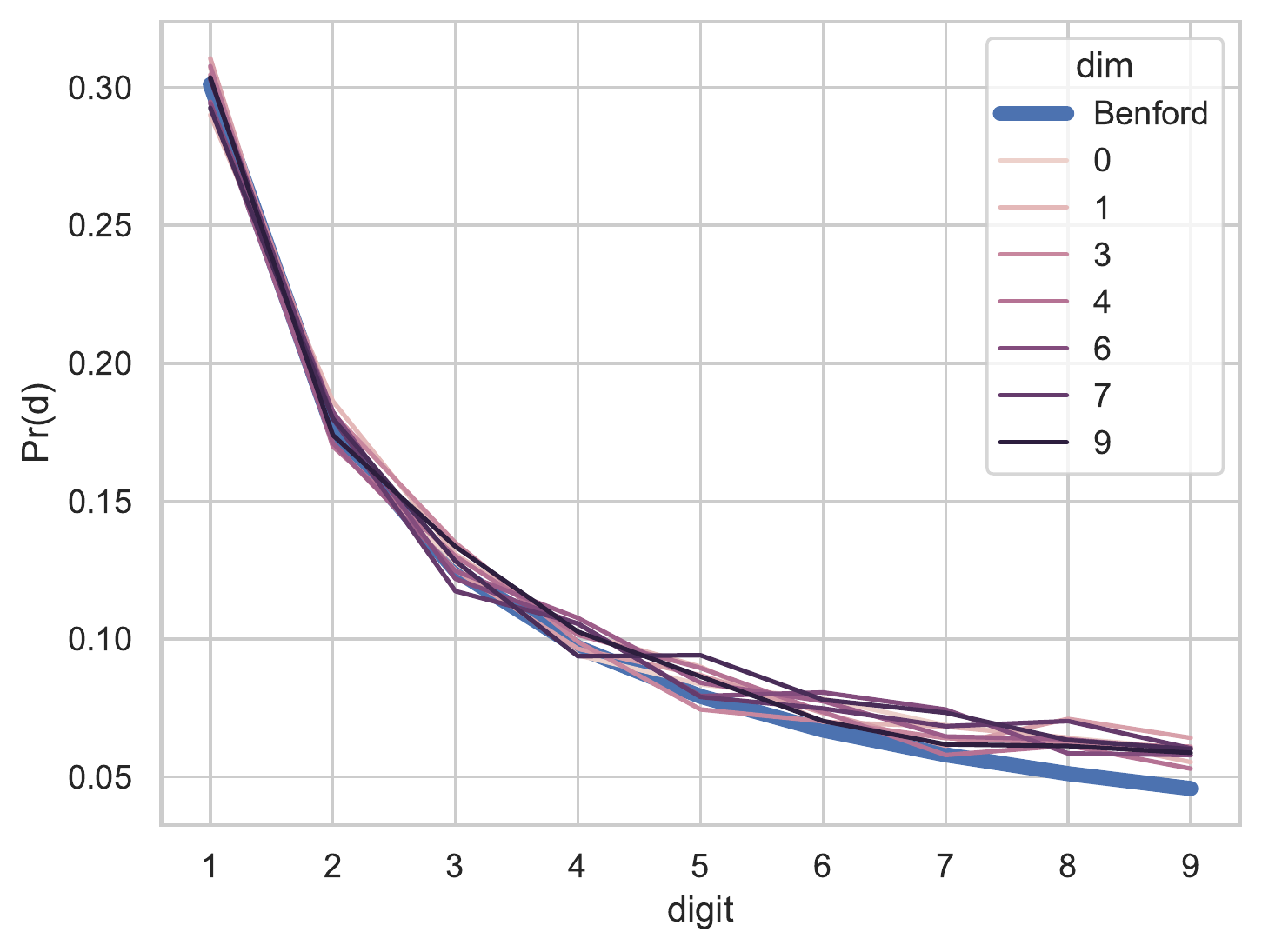}}
~   
\subfloat[Sobol']{
\includegraphics[width=0.5\linewidth,height=\textheight,keepaspectratio]{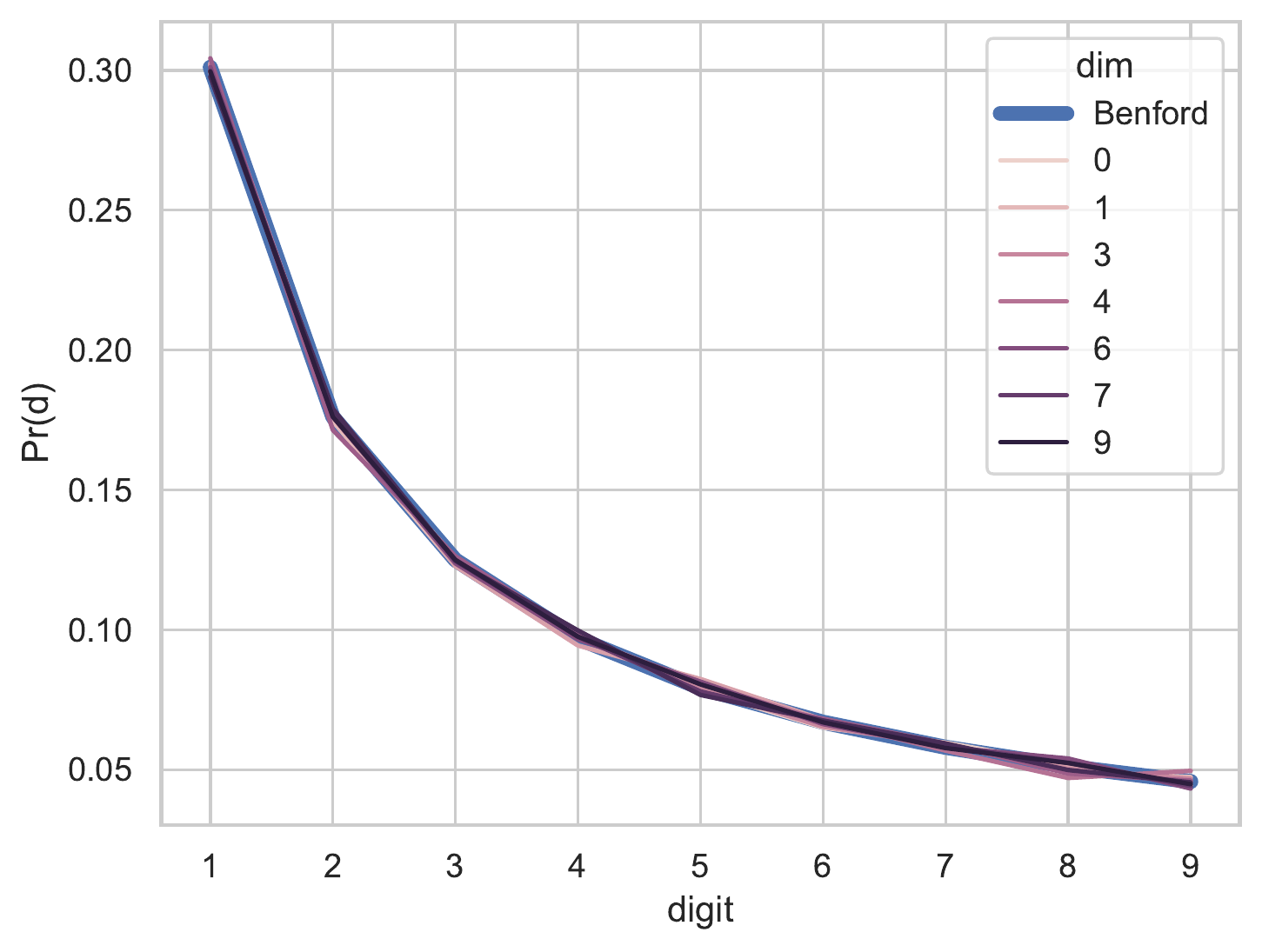}}

\subfloat[MC]{
\includegraphics[width=0.5\linewidth,height=\textheight,keepaspectratio]{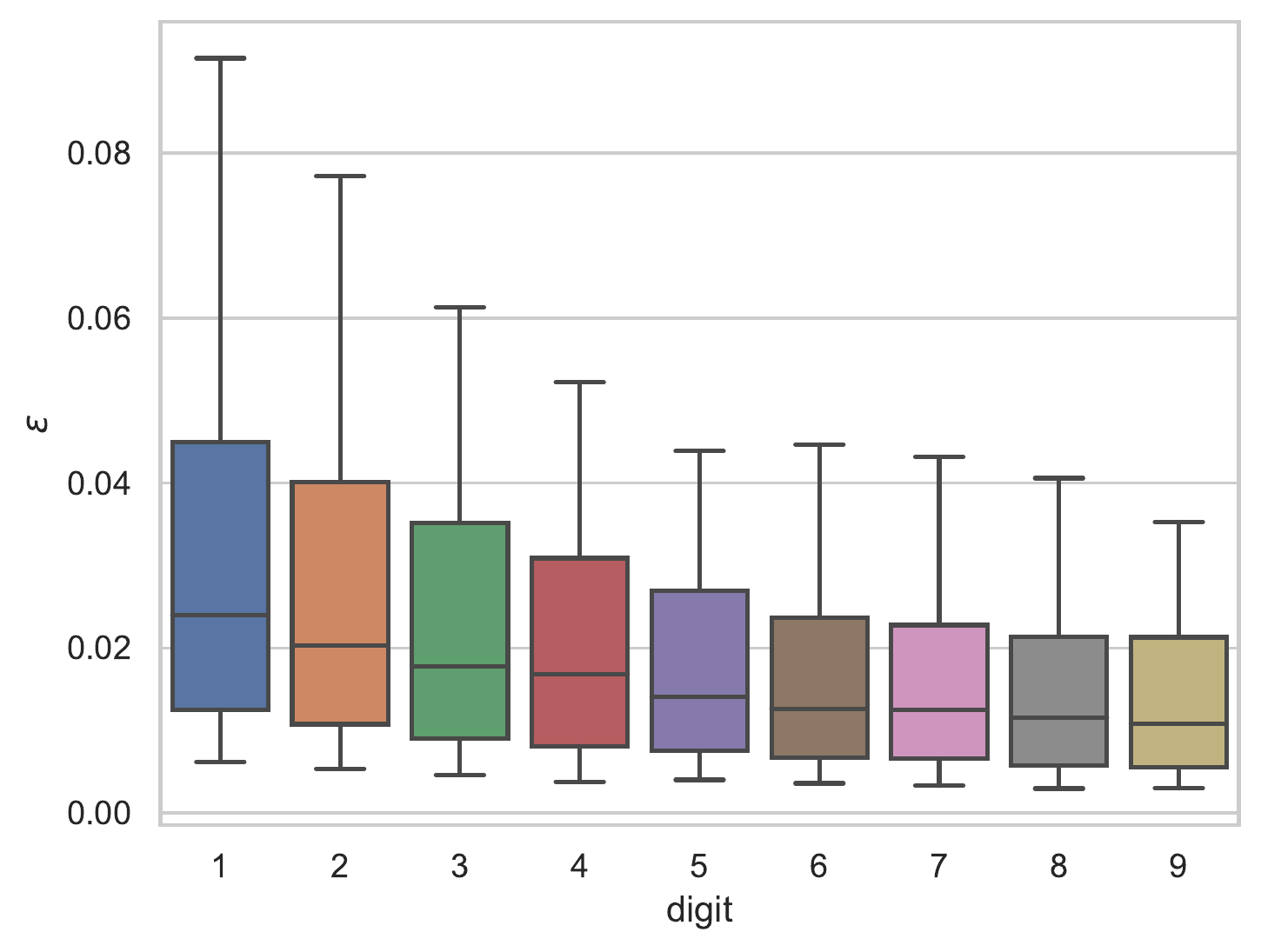}}
~   
\subfloat[Sobol']{
\includegraphics[width=0.5\linewidth,height=\textheight,keepaspectratio]{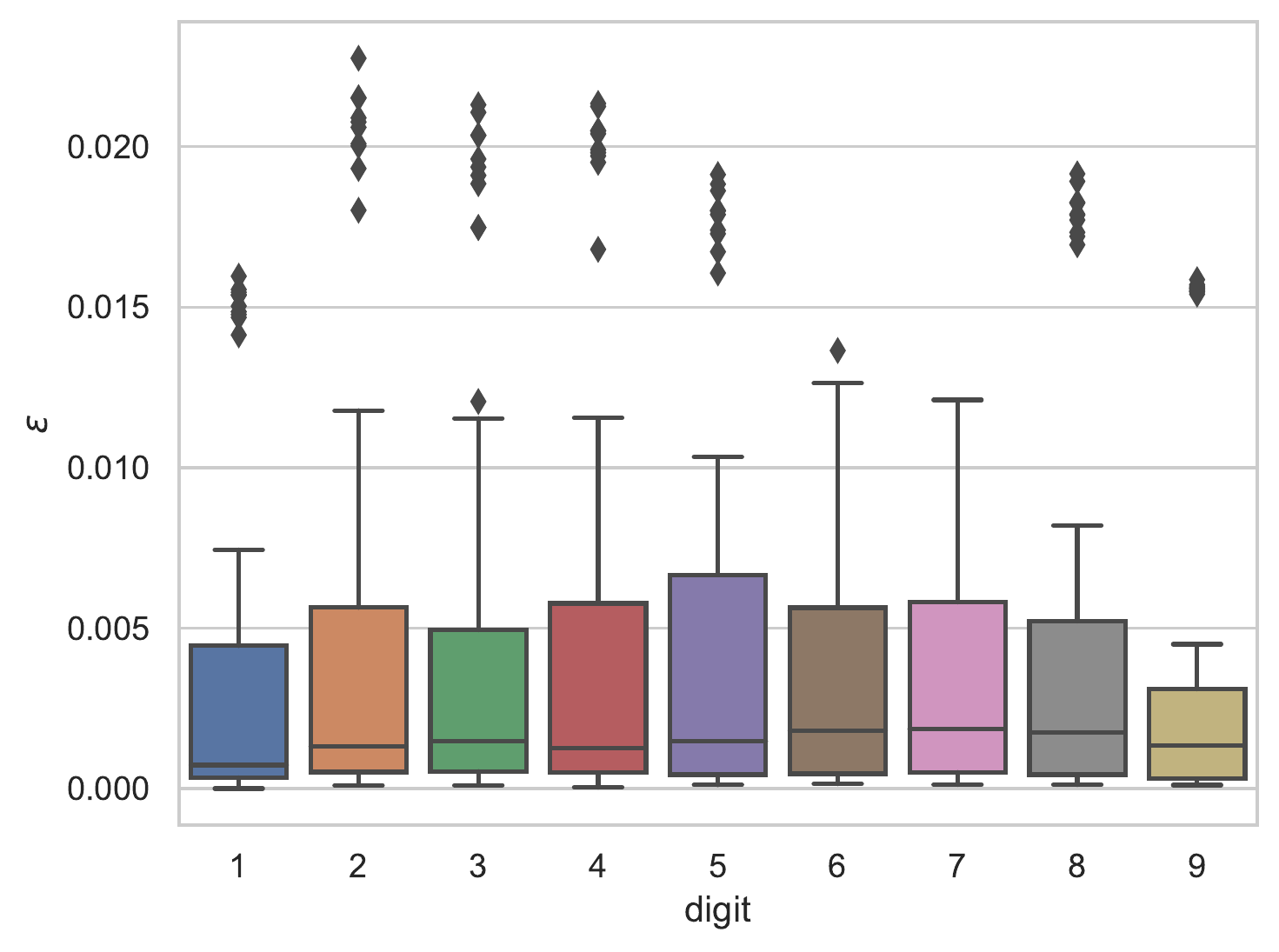}}

\caption{Conformance with the logarithmic law with respect to the number of dimensions for $N_s=32$ and $d=10$. Top subfigures show the logarithmic law and shades of purple represent the measure on the sample along a given dimension. Bottom subfigures show boxplot of the error.}
\label{fig:boxplot}
\end{figure}

Let's first take a look at~\cref{fig:boxplot}. It presents the conformance with Newcomb-Benford's law of two sets of $N_s=32, d=10$. The subfigures at the top present raw values whereas subfigures at the bottom present a boxplot of the NBD. Sobol' method clearly follows the best the logarithmic law even at a low number of sample---along all 10 dimensions. This is confirmed when looking at the error levels as there is an order of magnitude between the two. The boxplot shows something more interesting, there is a clear heteroscedasticity of the error in case of MC whereas Sobol' shows an homoscedasticity of the error. This indicates two things: \emph{(i)} from one run to another, Sobol' would produce more consistent samples in terms of quality; \emph{(ii)} it does not introduce biases as to favour more digits over the others.

\begin{figure}[!h]               
\centering
\subfloat[Flattened array]{
\includegraphics[width=0.5\linewidth,height=\textheight,keepaspectratio]{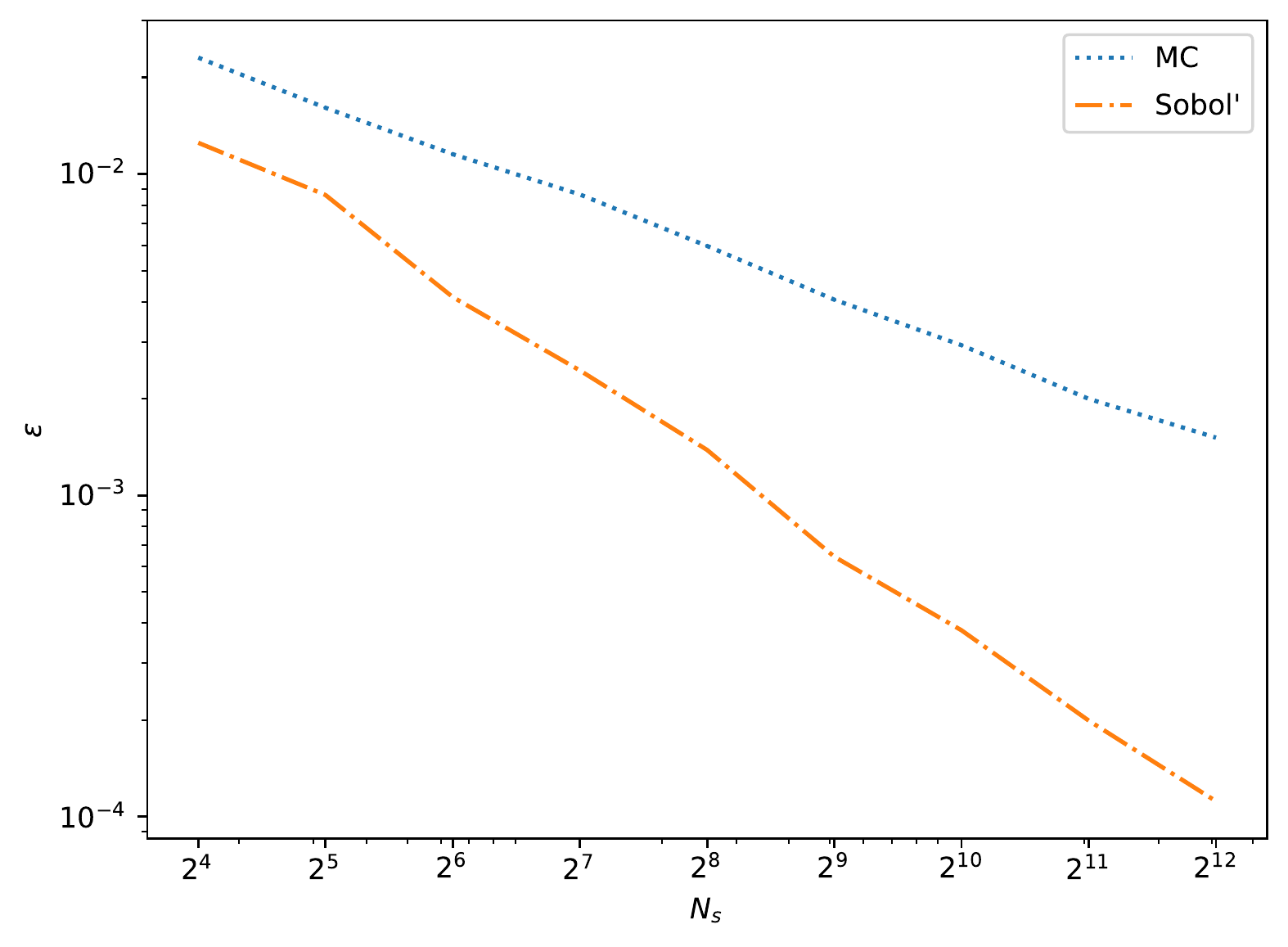}}
~
\subfloat[2D subprojections]{
\includegraphics[width=0.5\linewidth,height=\textheight,keepaspectratio]{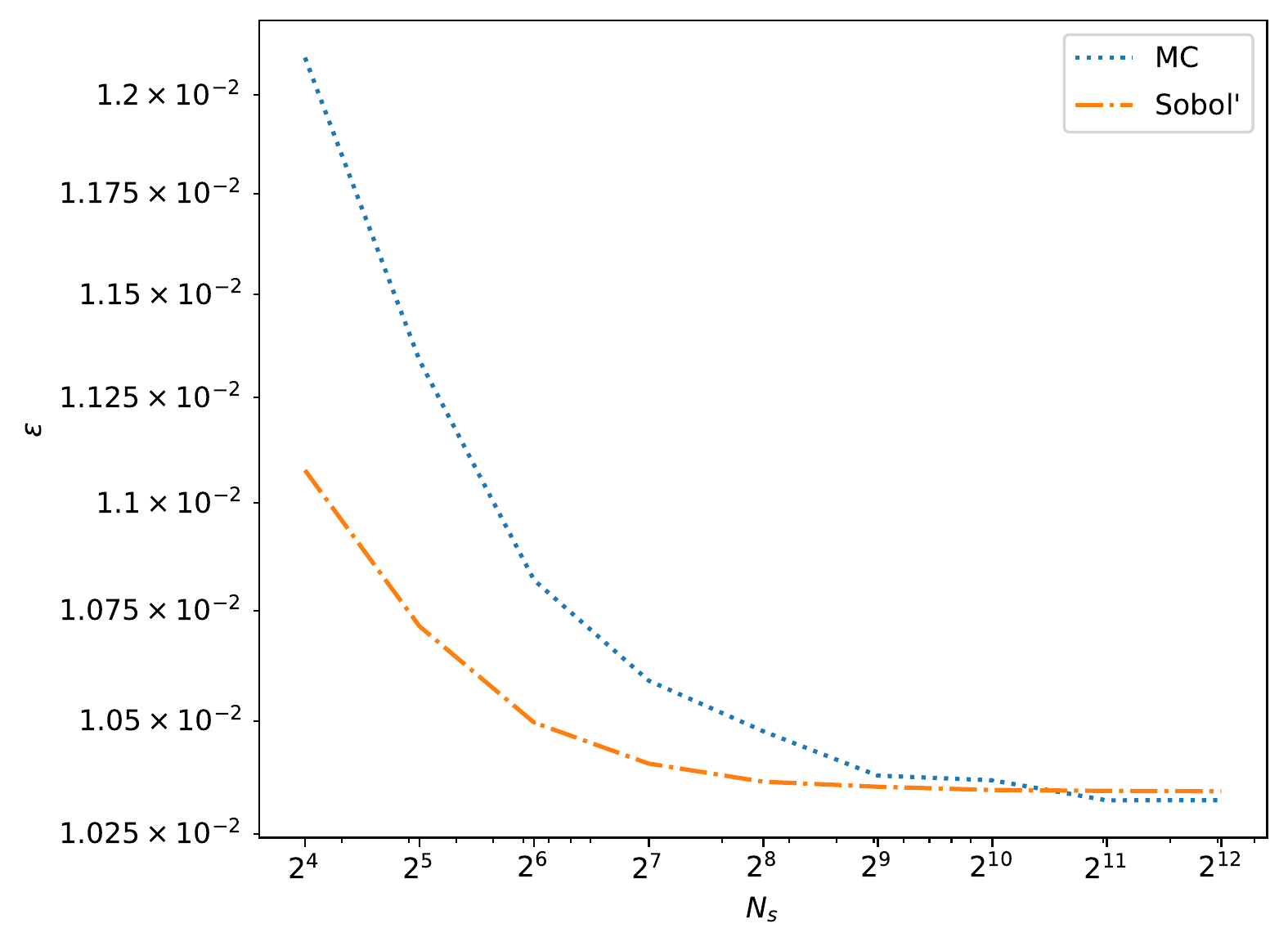}}

\caption{Convergence of the \emph{NB-discrepancy} with respect to $N_s$ for $d=10$.}
\label{fig:metric}
\end{figure}

Then, the convergence of NBD is assessed with respect to $N_s$ for $d=10$ in~\cref{fig:metric}. As expected, NBD converges with $N_s$ and at a faster rate and with lower values for Sobol' than MC. This is in accordance with classical convergence results found in~\citep{Kucherenko2015}. The joint-NBD error seems to saturate as $N_s$ increases.

Looking back at the case in~\cref{fig:lhs}, the flattened NBD metric gives, as expected, the same value of $0.0693$ for both designs. Whereas the 2-dimensional joint version gives $0.0321$ and $0.0406$---resp. for the design of the left and the design of the right. This is in accordance with results from the $C^2$-discrepancy which gives, resp., $0.0081$ and $0.0105$. The hierarchy between the design is correct using both methods. But as seen in~\cref{sec:method}, the numerical complexity of the joint-NBD is higher. Hence, if it is known that the designs to compare would not consist of permutations of coordinates (which is not the case for common LHS optimization methods for instance), the most efficient way can be used. Also, with large $N_s$ it seems that the joint version would not be able to be used to discriminate correctly the designs.

Note that both analyses were repeated with various number of samples---ranging from $N_s=32$ to $N_s=4096$---, and dimensions---ranging from $d=2$ to $d=100$. In all cases, the results agreed with each other.

\cleardoublepage
\section{Conclusion}\label{sec:conclusion}

This work proposes a new method to assess the uniformity of a design. This method is based on Newcomb-Benford's law, referred to as NB-discrepancy (NBD). The RMSE between the logarithmic law is compared with the empirical probabilities of a given sample which leads to a sensible metric. It has also been shown that this strategy can be applied to joint distributions of digits in order to assess the uniformity over the different dimensions of the design. NBD allows having a fast measure of uniformity with a numerical complexity of $\mathcal{O}(N_s \times d)$. In case the designs to compare would consist of permuted coordinates, such as during the optimization of LHS, the joint distribution of digits can be used in the same way. In both cases, the convergence properties have been shown with respect to the number of samples and dimensions.

Compared to classical discrepancy such as the $C^2$-discrepancy, it is noticeably easier to implement. In~\citep{Fang2006}, update strategies are given to update a LHS design and it leads to the same numerical complexity of NBD. But the individual operations are far more complex/expensive than simply counting the leading digits. Also, their update strategy is specific to the case of optimizing a LHS by permuting coordinates. The proposed method, on the other hand, is completely independent of the sampling strategy.

Being able to characterize a design is paramount as it determines the quality of the task it is used for. The proposed method provides an alternative to classical discrepancy methods used to control such design. It can be used to compare quantitatively designs, it is simple to implement and scalable which paves the way towards high-dimensional optimizations.


\section*{Acknowledgements}
The author acknowledges Prof. Art B. Owen from Stanford University for helpful discussions.

\bibliography{paper}

\end{document}